\documentclass[twocolumn,amsmath,amssymb]{revtex4}

\usepackage{graphicx}
\usepackage{dcolumn}
\usepackage{bm}

\begin{document}

\title{Hydrodynamic covariant symplectic structure from  bilinear Hamiltonian functions}

\author{Salvatore Capozziello}
 \email{capozziello@sa.infn.it}
 \affiliation{%
Dipartimento di Scienze Fisiche, Universit\'a di Napoli and INFN Sez. di Napoli,\\
Via Cinthia,  I-80126 Napoli
}%

\author{Salvatore De Martino}%
 \email{demartino@sa.infn.it}

\author{Stephan I. Tzenov}
 \email{tzenov@sa.infn.it}

\affiliation{%
Dipartimento di Fisica "E.R. Caianiello", Universit\'a di Salerno and INFN Sez. di Napoli\\
Gruppo Collegato di Salerno, Via S. Allende, I-84081 Baronissi
(SA)}%

\date{\today}

\begin{abstract}
Starting from generic bilinear Hamiltonians, constructed by
covariant vector, bivector or tensor fields, it is possible to
derive a general symplectic structure which leads to holonomic and
anholonomic formulations of Hamilton equations of motion directly
related to a hydrodynamic picture. This feature is gauge free and
it seems a deep link common to all interactions, electromagnetism
and gravity included. This scheme could lead toward a full
canonical quantization.
\end{abstract}


\maketitle

\vspace{10. mm} KEY WORDS:   Symplectic structure,  general
covariance, hydrodynamic equations, canonical Hamiltonian
mechanics.

\section{\label{Intro}Introduction}

It is well known that a self-consistent quantum field theory of
spacetime (quantum gravity) has not been achieved, up to now,
using standard quantization approaches. Specifically,  the request
of general coordinate invariance (one of the main features of
General Relativity) gives rise to unescapable  troubles in
understanding the dynamics of gravitational field. In fact, for a
physical (non-gravitational) field, one has to assign initially
the field amplitudes and their first time derivatives, in order to
determine the time development of such a field considered as a
dynamical entity. In General Relativity, these quantities are not
useful for dynamical determination since the metric field
$g_{\alpha \beta}$ can evolve at any time simply by a general
coordinate transformation. No change of physical observables is
the consequence of such an operation since it is nothing else but
a relabelling under which the theory is invariant. This apparent
"shortcoming" (from the quantum field theory point of view) means
that it is necessary a separation of metric degrees of freedom
into a part related to the true dynamical information and a part
related only to the coordinate system. From this viewpoint,
General Relativity is similar to classical Electromagnetism: the
coordinate invariance plays a role analogous to the
electromagnetic gauge invariance and in both cases (Lorentz and
gauge invariance) introduces redundant variables in order to
insure the maintenance of transformation properties. However,
difficulties come out as soon as one try to disentangle dynamical
from gauge variables. This operation is extremely clear in
Electromagnetism while it is not in General Relativity due to its
intrinsic non-linearity. A determination of independent dynamical
modes of gravitational field can be achieved when the theory is
cast into a canonical form involving the minimal number of degrees
of freedom which specify the state of the system. The canonical
formalism is essential in quantization program since it leads
directly to Poisson bracket relations among conjugate variables.
In order to realize it in any fundamental theory, one needs first
order field equations in time derivatives (Hamilton-like
equations) and a $(3+1)$-form of dynamics where time has been
unambiguously singled out. In General Relativity, the program has
been pursued using the first order Palatini approach
\cite{palatini}, where metric $g_{\alpha \beta}$ is taken into
account independently of affinity connections
$\Gamma^{\gamma}_{\alpha \beta}$ (this fact gives rise to first
order field equations) and the so called ADM formalism \cite{adm}
where $(3+1)$-dimensional notation has led to the definition of
gravitational Hamiltonian and time as a conjugate pair of
variables. However, the genuine fundament of General Relativity,
the  covariance of all coordinates without the distinction among
space and time, is impaired and, despite of innumerable efforts,
the full quantization of gravity has not been achieved up to now.
The main problems are related to the lack of a well-definite
Hilbert space and a quantum concept of measure for $g_{\alpha
\beta}$. An extreme consequence of this lack of full quantization
for gravity could be related to the dynamical variables: very
likely, the true variables could not be directly related to metric
but to something else as, for example, the connection
$\Gamma^{\gamma}_{\alpha\beta}$. Despite of this lack, a covariant
symplectic structure can be identified also in the framework of
General Relativity and then also this theory could be equipped
with the same features of other fundamental theories. This
statement does not still mean that the identification of a
symplectic structure immediately leads to a full quantization but
it could be a useful hint toward it.

The aim of this paper is to show that a prominent role in the
identification of a covariant symplectic structure is played by
bilinear Hamiltonians which have to be conserved. In fact, taking
into account generic Hamiltonian invariants, constructed by
covariant vectors, bivectors or tensors, it is possible to show
that a symplectic structure can be achieved in any case. By
specifying the nature of such vector fields (or, in general,
tensor invariants), it gives rise to intrinsically symplectic
structure which is always related to Hamilton-like equations (and
a Hamilton-Jacobi-like approach is always found). This works for
curvature invariants, Maxwell theory and so on. In any case, the
only basic assumption is that conservation laws (in Hamiltonian
sense) have to be identified in the framework of the theory.

The layout of the paper is the following. In Sec.II, we give the
generalities on the  symplectic structure and the canonical
description of mechanics. Sec.III is devoted to the discussion of
 symplectic structures which are also generally covariant. We show
that a covariant analogue of Hamilton equations can be derived
from covariant vector (or tensor) fields in holonomic and
anholonomic coordinates. In Sec. IV,  the covariant symplectic
structure is casted into the hydrodynamic picture leading to the
recovery of the covariant Hamilton equations. Sec.V is devoted to
 applications, discussion and conclusions.

\section{Generalities on the Symplectic Structure and the Canonical description}
In order to construct every fundamental theory of physics, it is
worth selecting the symplectic structure of the manifold on which
such a theory is formulated. This goal is achieved if suitable
symplectic conjugate variables and even-dimensional vector spaces
are chosen. Furthermore, we need an antisymmetric, covariant
tensor which is non-degenerate.

We are dealing with a symplectic structure if the couple
\begin{equation}
{\left\{ {\mathbf{E_{2n}}}, {\mathbf{w}} \right\}}, \label{s1}
\end{equation}
is defined, where $\mathbf{E_{2n}}$ is a vector space and the
tensor $\mathbf{w}$ on $\mathbf{E_{2n}}$ associates scalar
functions to pairs of vectors, that is
\begin{equation}
{\left[ \mathbf{x}, \mathbf{y} \right]} = \mathbf{w} {\left(
\mathbf{x}, \mathbf{y} \right)}, \label{s2}
\end{equation}
which is the {\it antiscalar} product. Such an operation satisfies
the following properties
\begin{eqnarray}
&& {\left[ \mathbf{x}, \mathbf{y} \right]} = -{\left[ \mathbf{y},
\mathbf{x} \right]} \qquad\qquad\qquad \forall
\mathbf{x}, \mathbf{y} \in \mathbf{E_{2n}} \label{s3}\\
& & {\left[ \mathbf{x}, \mathbf{y} + \mathbf{z} \right]} = {\left[
\mathbf{x}, \mathbf{y} \right]} + {\left[ \mathbf{x}, \mathbf{z}
\right]} \qquad \forall \mathbf{x}, \mathbf{y}, \mathbf{z} \in \mathbf{E_{2n}}, \label{s4}\\
& & \mbox{a} {\left[ \mathbf{x}, \mathbf{y} \right]} = {\left[
\mbox{a} \mathbf{x}, \mathbf{y} \right]} \qquad\qquad\qquad
\forall \mbox{a} \in R, \;\; \mathbf{x}, \mathbf{y} \in \mathbf{E_{2n}} \label{s5}\\
& & {\left[ \mathbf{x}, \mathbf{y} \right]} = 0 \qquad\qquad\qquad
\forall y \in \mathbf{E_{2n}} \; \Rightarrow \; \mathbf{x} = 0 \label{s6}\\
& & {\left[ \mathbf{x}, {\left[ \mathbf{y}, \mathbf{z} \right]}
\right]} + {\left[ \mathbf{y}, {\left[ \mathbf{z}, \mathbf{x}
\right]} \right]} + {\left[ \mathbf{z}, {\left[ \mathbf{x},
\mathbf{y} \right]} \right]} = 0 \, \forall
\mathbf{x},\mathbf{y},\mathbf{z} \in \mathbf{E_{2n}} \label{s6'}
\end{eqnarray}
The last one is the Jacobi cyclic identity.

If $\{\mathbf{e}_i \}$ is a vector basis in $\mathbf{E_{2n}}$, the
antiscalar product is completely singled out by the matrix
elements
\begin{equation}
w_{ij} = {\left[ \mathbf{e}_i, \mathbf{e}_j \right]}, \label{s7}
\end{equation}
where $\mathbf{w}$ is an antisymmetric matrix with determinant
different from zero. Every antiscalar product between two vectors
can be expressed as
\begin{equation}
{\left[ \mathbf{x}, \mathbf{y} \right]} = \mbox{\it w}_{ij}
\mbox{\it x}^i \mbox{\it y}^j, \label{s8}
\end{equation}
where $x^i$ and $y^j$ are the vector components in the given
basis.

The form of the matrix $\mathbf{w}$ and the relation (\ref{s8})
become considerably simpler if a canonical basis is taken into
account for $\mathbf{w}$. Since $\mathbf{w}$ is an antisymmetric
non-degenerate tensor, it is always possible to represent it
through the matrix
\begin{equation}
J=\left(
\begin{array}{cc}
0 & I \\ -I & 0
\end{array}
\right), \label{s9}
\end{equation}
where $I$ is a $(n \times n)$ unit matrix. Every basis where
$\mathbf{w}$ can be represented through the form (\ref{s9}) is a
{\it symplectic basis}. In other words, the symplectic bases are
the canonical bases for any antisymmetric non-degenerate tensor
$\mathbf{w}$ and can be characterized by the following conditions:
\begin{equation}
{\left[ \mathbf{e}_i, \mathbf{e}_j \right]} = 0,\qquad {\left[
\mathbf{e}_{n+i}, \mathbf{e}_{n+j} \right]} = 0,\qquad {\left[
\mathbf{e}_{i}, \mathbf{e}_{n+j} \right]} = \delta_{ij},
\label{s10}
\end{equation}
which have to be verified for every pair of values $i$ and $j$
ranging from 1 to $n$.

Finally, the expression of the antiscalar product between two
vectors, in a symplectic basis, is
\begin{equation}
{\left[ \mathbf{x}, \mathbf{y} \right]} = \sum \limits_{i=1}^{n}
{\left( \mbox{\it x}^{n+i} \mbox{\it y}^{i} - \mbox{\it x}^{i}
\mbox{\it y}^{n+i} \right)}, \label{s11}
\end{equation}
and a symplectic transformation in $\mathbf{E_{2n}}$ leaves
invariant the antiscalar product

\begin{equation}
\mathbf{S} {\left[ \mathbf{x}, \mathbf{y} \right]} = {\left[
\mathbf{S} (\mathbf{x}), \mathbf{S} (\mathbf{y}) \right]} =
{\left[ x, y \right]}. \label{s12}
\end{equation}
It is easy to see that standard Quantum Mechanics satisfies such
properties and so it is endowed with a symplectic structure.

On the other hand a standard canonical description can be sketched
as follows. For example, the relativistic Lagrangian of a charged
particle interacting with a vector field $A {\left( q; s \right)}$
is
\begin{equation}
{\cal L} {\left( q, u; s \right)} = {\frac {m u^2} {2}} - e u
\cdot A {\left( q; s \right)}, \label{Lagrangian}
\end{equation}
where the scalar product is defined as
\begin{equation}
z \cdot w = z_{\mu} w^{\mu} = \eta_{\mu \nu} z^{\mu} w^{\nu},
\label{Scalarprod}
\end{equation}
and the signature of the Minkowski spacetime is the usual one with
\begin{equation}
z_{\mu} = \eta_{\mu \nu} z^{\nu}, \qquad \qquad {\widehat{\eta}} =
{\rm diag} {\left( 1, -1, -1, -1 \right)}. \label{Signature}
\end{equation}
Furthermore, the contravariant vector $u^{\mu}$ with components $u
= {\left( u^0, u^1, u^2, u^3 \right)}$ is the four-velocity
\begin{equation}
u^{\mu} = {\frac {{\rm d} q^{\mu}} {{\rm d} s}}. \label{Fourveloc}
\end{equation}
The canonical conjugate momentum $\pi^{\mu}$ is defined as
\begin{equation}
\pi^{\mu} = \eta^{\mu \nu} {\frac {\partial {\cal L}} {\partial
u^{\nu}}} = m u^{\mu} - e A^{\mu}, \label{Canmomentum}
\end{equation}
so that the relativistic Hamiltonian can be written in the form
\begin{equation}
{\cal H} {\left( q, \pi; s \right)} = \pi \cdot u - {\cal L}
{\left( q, u; s \right)}. \label{Hamiltonia}
\end{equation}
Suppose now that we wish to use any other coordinate system
$x^{\alpha}$ as Cartesian, curvilinear, accelerated or rotating
one. Then the coordinates $q^{\mu}$ are functions of the
$x^{\alpha}$, which can be written explicitly as
\begin{equation}
q^{\mu} = q^{\mu} {\left( x^{\alpha} \right)}. \label{Newcoordin}
\end{equation}
The four-vector of particle velocity $u^{\mu}$ is transformed
according to the expression
\begin{equation}
u^{\mu} = {\frac {\partial q^{\mu}} {\partial x^{\alpha}}} {\frac
{{\rm d} x^{\alpha}} {{\rm d} s}} = {\frac {\partial q^{\mu}}
{\partial x^{\alpha}}} v^{\alpha}, \label{Transfourveloc}
\end{equation}
where
\begin{equation}
v^{\mu} = {\frac {{\rm d} x^{\mu}} {{\rm d} s}}.
\label{Fourvelocn}
\end{equation}
is the transformed four-velocity expressed in terms of the new
coordinates. The vector field $A^{\mu}$ is also transformed as a
vector
\begin{equation}
{\cal A}^{\mu} = {\frac {\partial x^{\mu}} {\partial q^{\alpha}}}
A^{\alpha}. \label{Transvecfield}
\end{equation}
In the new coordinate system $x^{\alpha}$ the Lagrangian
(\ref{Lagrangian}) becomes
\begin{equation}
{\cal L} {\left( x, v; s \right)} = g_{\mu \nu} {\left[ {\frac {m}
{2}} v^{\mu} v^{\nu} - e v^{\mu} {\cal A}^{\nu} {\left( x; s
\right)} \right]}, \label{Lagrangianncs}
\end{equation}
where
\begin{equation}
g_{\alpha \beta} = \eta_{\mu \nu} {\frac {\partial q^{\mu}}
{\partial x^{\alpha}}} {\frac {\partial q^{\nu}} {\partial
x^{\beta}}}. \label{Metricdef}
\end{equation}
The Lagrange equations can be written in the usual form
\begin{equation}
{\frac {{\rm d}} {{\rm d} s}} {\left( {\frac {\partial {\cal L}}
{\partial v^{\lambda}}} \right)} - {\frac {\partial {\cal L}}
{\partial x^{\lambda}}} = 0. \label{Lagrangeequ}
\end{equation}
In the case of a free particle (no interaction with an external
vector field), we have
\begin{equation}
{\frac {{\rm d}} {{\rm d} s}} {\left( g_{\lambda \mu} v^{\mu}
\right)} - {\frac {1} {2}} {\frac {\partial g_{\mu \nu}} {\partial
x^{\lambda}}} v^{\mu} v^{\nu} = 0. \label{Lagrangeequ1}
\end{equation}
Specifying the covariant velocity $v_{\lambda}$ as
\begin{equation}
v_{\lambda} = g_{\lambda \mu} v^{\mu}, \label{Covariantvel}
\end{equation}
and using the well-known identity for connections
$\Gamma^{\alpha}_{\mu\nu}$
\begin{equation}
{\frac {\partial g_{\mu \nu}} {\partial x^{\lambda}}} =
{\Gamma}^{\alpha}_{\lambda \mu} g_{\alpha \nu} +
{\Gamma}^{\alpha}_{\lambda \nu} g_{\alpha \mu}, \label{Identity}
\end{equation}
we obtain
\begin{equation}
{\frac {{\rm D} v_{\lambda}} {{\rm D} s}} = {\frac {{\rm d}
v_{\lambda}} {{\rm d} s}} - {\Gamma}^{\mu}_{\lambda \nu} v^{\nu}
v_{\mu} = 0. \label{Covardiff}
\end{equation}
Here ${\rm D} v_{\lambda} / {\rm D} s$ denotes the covariant
derivative of the covariant velocity $v_{\lambda}$ along the curve
$x^{\nu} (s)$. Using Eqs. (\ref{Covariantvel}) and
(\ref{Identity}) and the fact that the affine connection
${\Gamma}^{\lambda}_{\mu \nu}$ is symmetric in the indices $\mu$
and $\nu$, we obtain the equation of motion for the contravariant
vector $v^{\lambda}$
\begin{equation}
{\frac {{\rm D} v^{\lambda}} {{\rm D} s}} = {\frac {{\rm d}
v^{\lambda}} {{\rm d} s}} + {\Gamma}^{\lambda}_{\mu \nu} v^{\mu}
v^{\nu} = 0. \label{Geodesicequ}
\end{equation}

Before we pass over to the Hamiltonian description, let us note
that the generalized momentum $p_{\mu}$ is defined as
\begin{equation}
p_{\mu} = {\frac {\partial {\cal L}} {\partial v^{\mu}}} = m
g_{\mu \nu} v^{\nu}, \label{Genmomentum}
\end{equation}
while, from Lagrange equations of motion, we obtain
\begin{equation}
{\frac {{\rm d} p_{\mu}} {{\rm d} s}} = {\frac {\partial {\cal L}}
{\partial x^{\mu}}}. \label{Lagraeqmot}
\end{equation}

The transformation from ${\left( x^{\mu}, v^{\mu}; s \right)}$ to
${\left( x^{\mu}, p_{\mu}; s \right)}$ can be accomplished by
means of a Legendre transformation, and instead of the Lagrangian
(\ref{Lagrangianncs}), we consider the Hamilton function
\begin{equation}
{\cal H} {\left( x, p; s \right)} = p_{\mu} v^{\mu} - {\cal L}
{\left( x, v; s \right)}. \label{Hamiltonian}
\end{equation}
The differential of the Hamiltonian in terms of $x$, $p$ and $s$
is given by
\begin{equation}
{\rm d} {\cal H} = {\frac {\partial {\cal H}} {\partial x^{\mu}}}
{\rm d} x^{\mu} + {\frac {\partial {\cal H}} {\partial p_{\mu}}}
{\rm d} p_{\mu} + {\frac {\partial {\cal H}} {\partial s}} {\rm d}
s. \label{Differham}
\end{equation}
On the other hand, from Eq.(\ref{Hamiltonian}), we have
\begin{equation}
{\rm d} {\cal H} = v^{\mu} {\rm d} p_{\mu} + p_{\mu} {\rm d}
v^{\mu} - {\frac {\partial {\cal L}} {\partial v^{\mu}}} {\rm d}
v^{\mu} - {\frac {\partial {\cal L}} {\partial x^{\mu}}} {\rm d}
x^{\mu} - {\frac {\partial {\cal L}} {\partial s}} {\rm d} s.
\label{Differhamil}
\end{equation}
Taking into account the defining Eq.(\ref{Genmomentum}), the
second and the third term on the right-hand-side of
Eq.(\ref{Differhamil}) cancel out. Eq.(\ref{Lagraeqmot}) can be
further used to cast Eq.(\ref{Differhamil}) into the form
\begin{equation}
{\rm d} {\cal H} = v^{\mu} {\rm d} p_{\mu} - {\frac {{\rm d}
p_{\mu}} {{\rm d} s}} {\rm d} x^{\mu} - {\frac {\partial {\cal L}}
{\partial s}} {\rm d} s, \label{Differhamilint}
\end{equation}
Comparison between Eqs.(\ref{Differham}) and
(\ref{Differhamilint}) yields the Hamilton equations of motion
\begin{equation}
{\frac {{\rm d} x^{\mu}} {{\rm d} s}} = {\frac {\partial {\cal H}}
{\partial p_{\mu}}}, \qquad \qquad {\frac {{\rm d} p_{\mu}} {{\rm
d} s}} = - {\frac {\partial {\cal H}} {\partial x^{\mu}}},
\label{Hamiltoneqmot}
\end{equation}
where the Hamiltonian is given by
\begin{equation}
{\cal H} {\left( x, p; s \right)} = {\frac {g^{\mu \nu}} {2m}}
p_{\mu} p_{\nu} + {\frac {e} {m}} p_{\mu} {\cal A}^{\mu}.
\label{Hamiltonianf}
\end{equation}
In the case of a free particle,  the Hamilton equations can be
written explicitly as
\begin{equation}
{\frac {{\rm d} x^{\mu}} {{\rm d} s}} = {\frac {g^{\mu \nu}} {m}}
p_{\nu}, \qquad \qquad {\frac {{\rm d} p_{\lambda}} {{\rm d} s}} =
- {\frac {1} {2m}} {\frac {\partial g^{\mu \nu}} {\partial
x^{\lambda}}} p_{\mu} p_{\nu}. \label{Hamileqmot}
\end{equation}
To obtain the equations of motion we need the expression
\begin{equation}
{\frac {\partial g^{\mu \nu}} {\partial x^{\lambda}}} = -
{\Gamma}^{\mu}_{\lambda \alpha} g^{\alpha \nu} -
{\Gamma}^{\nu}_{\lambda \alpha} g^{\alpha \mu}, \label{Identity1}
\end{equation}
which can be derived from the obvious identity
\begin{equation}
{\frac {\partial} {\partial x^{\lambda}}} {\left( g^{\mu \alpha}
g_{\alpha \nu} \right)} = 0, \label{Identity2}
\end{equation}
and Eq.(\ref{Identity}). From the second of Eqs.
(\ref{Hamileqmot}), we obtain
\begin{equation}
{\frac {{\rm D} p_{\lambda}} {{\rm D} s}} = {\frac {{\rm d}
p_{\lambda}} {{\rm d} s}} - {\Gamma}^{\mu}_{\lambda \nu} v^{\nu}
p_{\mu} = 0, \label{Covardiff1}
\end{equation}
similar to equation (\ref{Covardiff}). Differentiating the first
of the Hamilton equations (\ref{Hamileqmot}) with respect to $s$
and taking into account equations (\ref{Identity1}) and
(\ref{Covardiff1}), we again arrive to the equation for the
geodesics (\ref{Geodesicequ}).

Let us now show that on a generic curved (torsion-free) manifolds
the Poisson brackets are conserved. To achieve this result, we
need the following identities
\begin{equation}
g^{\mu \nu} = g^{\nu \mu} = \eta^{\alpha \beta} {\frac {\partial
x^{\mu}} {\partial q^{\alpha}}} {\frac {\partial x^{\nu}}
{\partial q^{\beta}}}, \label{Append1}
\end{equation}
\begin{equation}
{\frac {\partial^2 x^{\lambda}} {\partial q^{\alpha} \partial
q^{\beta}}} = - \Gamma^{\lambda}_{\mu \nu} {\frac {\partial
x^{\mu}} {\partial q^{\alpha}}} {\frac {\partial x^{\nu}}
{\partial q^{\beta}}}, \label{Append2}
\end{equation}
$\blacksquare$ To prove (\ref{Append2}), we differentiate the
obvious identity
\begin{equation}
{\frac {\partial x^{\lambda}} {\partial q^{\rho}}} {\frac
{\partial q^{\rho}} {\partial x^{\nu}}} = \delta^{\lambda}_{\nu}.
\label{Identit}
\end{equation}
As a result, we find
\begin{equation}
\Gamma^{\lambda}_{\mu \nu} = {\frac {\partial x^{\lambda}}
{\partial q^{\rho}}} {\frac {\partial^2 q^{\rho}} {\partial
x^{\mu} \partial x^{\nu}}} = - {\frac {\partial q^{\rho}}
{\partial x^{\nu}}} {\frac {\partial q^{\sigma}} {\partial
x^{\mu}}} {\frac {\partial^2 x^{\lambda}} {\partial q^{\rho}
\partial q^{\sigma}}}. \label{Identit1}
\end{equation}
$\blacksquare$

The next step is to calculate the fundamental Poisson brackets in
terms of the variables ${\left( x^{\mu}, p_{\nu} \right)}$,
initially defined using the canonical variables ${\left( q^{\mu},
\pi_{\nu} \right)}$ according to the relation
\begin{equation}
{\left[ U, V \right]} = {\frac {\partial U} {\partial q^{\mu}}}
{\frac {\partial V} {\partial \pi_{\mu}}} - {\frac {\partial V}
{\partial q^{\mu}}} {\frac {\partial U} {\partial \pi_{\mu}}},
\label{Poisbra}
\end{equation}
where $U {\left( q^{\mu}, \pi_{\nu} \right)}$ and $V {\left(
q^{\mu}, \pi_{\nu} \right)}$ are arbitrary functions. Making use
of Eqs.(\ref{Canmomentum}) and (\ref{Transfourveloc}), we know
that the variables
\begin{equation}
q^{\mu} \quad \Leftrightarrow \quad \pi_{\mu} = m u_{\mu} = m
\eta_{\mu \nu} u^{\nu} = m \eta_{\mu \nu} {\frac {\partial
q^{\nu}} {\partial x^{\alpha}}} v^{\alpha}, \label{Flatcanvar}
\end{equation}
form a canonical conjugate pair. Using Eq.(\ref{Genmomentum}), we
would like to check whether the variables
\begin{equation}
x^{\mu} \quad \Leftrightarrow \quad p_{\mu} = m g_{\mu \nu}
v^{\nu} = g_{\mu \nu} \eta^{\alpha \lambda} \pi_{\lambda} {\frac
{\partial x^{\nu}} {\partial q^{\alpha}}}, \label{Curvecanvar}
\end{equation}
form a canonical conjugate pair. We have
\begin{eqnarray}
{\left[ U, V \right]} = {\left[ {\frac {\partial U} {\partial
x^{\alpha}}} {\frac {\partial x^{\alpha}} {\partial q^{\mu}}} +
{\frac {\partial U} {\partial p_{\sigma}}} \eta^{\beta \lambda}
\pi_{\lambda} {\frac {\partial} {\partial q^{\mu}}} {\left(
g_{\sigma \nu} {\frac {\partial x^{\nu}} {\partial q^{\beta}}}
\right)} \right]} \nonumber \\
\times {\frac {\partial V} {\partial p_{\alpha}}} g_{\alpha \chi}
\eta^{\rho \mu} {\frac {\partial x^{\chi}} {\partial q^{\rho}}}
\nonumber \\
- {\left[ {\frac {\partial V} {\partial x^{\alpha}}} {\frac
{\partial x^{\alpha}} {\partial q^{\mu}}} + {\frac {\partial V}
{\partial p_{\sigma}}} \eta^{\beta \lambda} \pi_{\lambda} {\frac
{\partial} {\partial q^{\mu}}} {\left( g_{\sigma \nu} {\frac
{\partial x^{\nu}} {\partial q^{\beta}}}
\right)} \right]} \nonumber \\
\times {\frac {\partial U} {\partial p_{\alpha}}} g_{\alpha \chi}
\eta^{\rho \mu} {\frac {\partial x^{\chi}} {\partial q^{\rho}}}.
\label{Poisbrcal}
\end{eqnarray}
The first and the third term on the right-hand-side of
Eq.(\ref{Poisbrcal}) can be similarly manipulated as follows
\begin{eqnarray}
{\rm {\bf I}-st} \; \; {\rm term} = {\frac {\partial U} {\partial
x^{\alpha}}} {\frac {\partial V} {\partial p_{\beta}}} g_{\beta
\chi} \eta^{\rho \mu} {\frac {\partial x^{\chi}} {\partial
q^{\rho}}} {\frac {\partial x^{\alpha}} {\partial q^{\mu}}}
\nonumber \\
= g_{\beta \chi} g^{\chi \alpha} {\frac {\partial U} {\partial
x^{\alpha}}} {\frac {\partial V} {\partial p_{\beta}}} = {\frac
{\partial U} {\partial x^{\alpha}}} {\frac {\partial V} {\partial
p_{\alpha}}}, \label{Firthiterm}
\end{eqnarray}
\begin{equation}
{\rm {\bf III}-rd} \; \; {\rm term} = - {\frac {\partial V}
{\partial x^{\alpha}}} {\frac {\partial U} {\partial p_{\alpha}}}.
\label{Firtterm}
\end{equation}
Next, we manipulate the second term on the right-hand-side of
Eq.(\ref{Poisbrcal}). We obtain
\begin{eqnarray}
{\rm {\bf II}-nd} \; \; {\rm term} = {\frac {\partial U} {\partial
p_{\sigma}}} {\frac {\partial V} {\partial p_{\alpha}}} g_{\alpha
\chi} \eta^{\rho \mu} {\frac {\partial x^{\chi}} {\partial
q^{\rho}}} \eta^{\beta \lambda} \pi_{\lambda} \nonumber \\
\times {\left[ g_{\sigma \nu} {\frac {\partial^2 x^{\nu}}
{\partial q^{\mu} \partial q^{\beta}}} + {\frac {\partial x^{\nu}}
{\partial q^{\beta}}} {\frac {\partial g_{\sigma \nu}} {\partial
x^{\gamma}}} {\frac {\partial x^{\gamma}} {\partial q^{\mu}}}
\right]} \nonumber \\
= {\frac {\partial U} {\partial p_{\sigma}}} {\frac {\partial V}
{\partial p_{\alpha}}} g_{\alpha \chi} \eta^{\rho \mu} {\frac
{\partial x^{\chi}} {\partial q^{\rho}}} \eta^{\beta \lambda}
\pi_{\lambda} \nonumber \\
\times {\left[ -g_{\sigma \nu} \Gamma^{\nu}_{\gamma \delta} {\frac
{\partial x^{\gamma}} {\partial q^{\mu}}} {\frac {\partial
x^{\delta}} {\partial q^{\beta}}} + {\frac {\partial x^{\delta}}
{\partial q^{\beta}}} {\frac {\partial x^{\gamma}} {\partial
q^{\mu}}} {\left( \Gamma^{\nu}_{\gamma \sigma} g_{\nu \delta} +
\Gamma^{\nu}_{\gamma \delta} g_{\nu \sigma} \right)} \right]}
\nonumber \\
= {\frac {\partial U} {\partial p_{\sigma}}} {\frac {\partial V}
{\partial p_{\alpha}}} g_{\alpha \chi} g^{\chi \gamma} \eta^{\beta
\lambda} \pi_{\lambda} {\frac {\partial x^{\delta}} {\partial
q^{\beta}}} g_{\nu \delta} \Gamma^{\nu}_{\gamma \sigma} \nonumber
\\
= {\frac {\partial U} {\partial p_{\sigma}}} {\frac {\partial V}
{\partial p_{\beta}}} g_{\mu \nu} \eta^{\alpha \lambda}
\pi_{\lambda} {\frac {\partial x^{\nu}} {\partial q^{\alpha}}}
\Gamma^{\mu}_{\beta \sigma} \nonumber
\end{eqnarray}
\begin{equation}
= \Gamma^{\lambda}_{\mu \nu} p_{\lambda} {\frac {\partial U}
{\partial p_{\nu}}} {\frac {\partial V} {\partial p_{\mu}}}.
\label{Secondterm}
\end{equation}
The fourth term is similar to the second one but with $U$ and $V$
interchanged
\begin{equation}
{\rm {\bf IV}-th} \; \; {\rm term} = - \Gamma^{\lambda}_{\mu \nu}
p_{\lambda} {\frac {\partial U} {\partial p_{\mu}}} {\frac
{\partial V} {\partial p_{\nu}}}. \label{Fourterm}
\end{equation}
In the absence of torsion, the affine connection
$\Gamma^{\lambda}_{\mu \nu}$ is symmetric with respect to the
lower indices, so that the second and the fourth term on the
right-hand-side of Eq.(\ref{Poisbrcal}) cancel each other.
Therefore,
\begin{equation}
{\left[ U, V \right]} = {\frac {\partial U} {\partial x^{\mu}}}
{\frac {\partial V} {\partial p_{\mu}}} - {\frac {\partial V}
{\partial x^{\mu}}} {\frac {\partial U} {\partial p_{\mu}}},
\label{Poisbracon}
\end{equation}
which means that the fundamental Poisson brackets are conserved.
On the other hand, this implies that the variables ${\left\{
x^{\mu}, p_{\nu} \right\}}$ are a canonical conjugate pair.

As a final remark, we have to say that considering a generic
metric $g_{\alpha\beta}$ and a  connection
$\Gamma^{\alpha}_{\mu\nu}$ is related to the fact that we are
passing from a Minkowski-flat spacetime (local inertial reference
frame) to an accelerated reference frame (curved spacetime). In
what follows, we want to show that a generic bilinear Hamiltonian
invariant, which is conformally conserved, gives always rise to a
canonical symplectic structure. The specific theory is assigned by
the vector (or tensor) fields which define the Hamiltonian
invariant.

\section{A symplectic structure compatible with general covariance }
The above considerations can be linked together leading to a more
general scheme where a covariant symplectic structure is achieved.
Summarizing, the main points which we need are: $i)$ an
even-dimensional vector space $\mathbf{E_{2n}}$ equipped with an
antiscalar product satisfying the algebra (\ref{s3})-(\ref{s6'});
$ii)$ generic vector fields defined on such a space which have to
satisfy the Poisson brackets; $iii)$ first-order equations of
motion which can be read as Hamilton-like equations; $iv)$ general
covariance which has to be preserved.

Such a program can be pursued by taking into account covariant and
contravariant vector fields. In fact,  it is possible to construct
the Hamiltonian invariant
\begin{equation}
{\cal H} = V^{\alpha} V_{\alpha}, \label{s26}
\end{equation}
which is a scalar quantity satisfying the relation
\begin{equation}
\delta {\cal H} = \delta {\left( V^{\alpha} V_{\alpha} \right)} =
0, \label{s27}
\end{equation}
being $\delta$ a spurious variation due to the transport. It is
worth stressing that the vectors $V^{\alpha}$ and $V_{\alpha}$ are
not specified and the following considerations are completely
general. Eq.(\ref{s26}) is a so called {\it "already
parameterized"} invariant which can constitute the "density" of a
parameterized action principle where the time coordinate is not
distinguished {\it a priori} from the other coordinates
\cite{lanczos,schwinger}.

Let us now take into account the intrinsic variation of
$V^{\alpha}$. On a generic curved manifold, we have
\begin{equation}
D V^{\alpha} = d V^{\alpha} - \delta V^{\alpha} = \partial_{\beta}
V^{\alpha} d x^{\beta} - \delta V^{\alpha}, \label{s28}
\end{equation}
where $D$ is the intrinsic variation, $d$ the total variation and
$\delta$ the spurious variation due to the transport on the curved
manifold.  The spurious variation has a very important meaning
since, in General Relativity, if such a variation for a given
quantity is equal to zero, this means that the quantity is
conserved. From the definition of covariant derivative, applied to
the contravariant vector, we have
\begin{equation}
D V^{\alpha} = \partial_{\beta} V^{\alpha} d x^{\beta} +
\Gamma^{\alpha}_{\sigma \beta} V^{\sigma} d x^{\beta}, \label{s29}
\end{equation}
and
\begin{equation}
\nabla_{\beta} V^{\alpha} = \partial_{\beta} V^{\alpha} +
\Gamma^{\alpha}_{\sigma \beta} V^{\sigma}, \label{s30}
\end{equation}
and then
\begin{equation}
\delta V^{\alpha} = - \Gamma^{\alpha}_{\sigma \beta} V^{\sigma} d
x^{\beta}. \label{s31}
\end{equation}
Analogously, for the covariant derivative applied to the covariant
vector,
\begin{equation}
D V_{\alpha} = d V_{\alpha} - \delta V_{\alpha} = \partial_{\beta}
V_{\alpha} d x^{\beta} - \delta V_{\alpha}, \label{s32}
\end{equation}
and then
\begin{equation}
D V_{\alpha} = \partial_{\beta} V_{\alpha} d x^{\beta} -
\Gamma^{\sigma}_{\alpha \beta} V_{\sigma} d x^{\beta}, \label{s33}
\end{equation}
and
\begin{equation}
\nabla_{\beta} V_{\alpha} = \partial_{\beta} V_{\alpha}  -
\Gamma^{\sigma}_{\alpha \beta} V_{\sigma}. \label{s34}
\end{equation}
The spurious variation is now
\begin{equation}
\delta V_{\alpha} = \Gamma^{\sigma}_{\alpha \beta} V_{\sigma} d
x^{\beta}. \label{s35}
\end{equation}
Developing the variation (\ref{s27}), we have
\begin{equation}
\delta {\cal H} = V_{\alpha} \delta V^{\alpha} + V^{\alpha} \delta
V_{\alpha}, \label{s36}
\end{equation}
and
\begin{equation}
{\frac {\delta {\cal H}} {d x^{\beta}}} = V_{\alpha} {\frac
{\delta V^{\alpha}} {d x^{\beta}}} + V^{\alpha} {\frac {\delta
V_{\alpha}} {d x^{\beta}}}, \label{s37}
\end{equation}
which becomes
\begin{equation}
{\frac {\delta {\cal H}} {d x^{\beta}}} = {\frac {\delta
V^{\alpha}} {d x^{\beta}}} {\frac {\partial {\cal H}} {\partial
V^{\alpha}}} + {\frac {\delta V_{\alpha}} {d x^{\beta}}} {\frac
{\partial {\cal H}} {\partial V_{\alpha}}}, \label{s38}
\end{equation}
being
\begin{equation}
{\frac {\partial {\cal H}} {\partial V^{\alpha}}} = V_{\alpha},
\qquad {\frac {\partial {\cal H}} {\partial V_{\alpha}}} =
V^{\alpha}. \label{s39}
\end{equation}
From  Eqs.(\ref{s31}) and (\ref{s35}), it is
\begin{equation}
{\frac {\delta V^{\alpha}} {d x^{\beta}}} = -
\Gamma^{\alpha}_{\sigma \beta} V^{\sigma} = -
\Gamma^{\alpha}_{\sigma \beta} {\left( {\frac {\partial {\cal H}}
{\partial V_{\sigma}}} \right)}, \label{s40}
\end{equation}
\begin{equation}
{\frac {\delta V_{\alpha}} {d x^{\beta}}} =
\Gamma^{\sigma}_{\alpha \beta} V_{\sigma} =
\Gamma^{\sigma}_{\alpha \beta} {\left( {\frac {\partial {\cal H}}
{\partial V^{\sigma}}} \right)}, \label{s41}
\end{equation}
and substituting into Eq.(\ref{s38}), we have
\begin{equation}
{\frac {\delta {\cal H}} {d x^{\beta}}} = -
\Gamma^{\alpha}_{\sigma \beta} {\left( {\frac {\partial {\cal H}}
{\partial V_{\sigma}}} \right)} {\left( {\frac {\partial {\cal H}}
{\partial V^{\alpha}}} \right)} + \Gamma^{\sigma}_{\alpha \beta}
{\left( {\frac {\partial {\cal H}} {\partial V_{\alpha}}} \right)}
{\left( {\frac {\partial {\cal H}} {\partial V^{\sigma}}}
\right)}, \label{s42}
\end{equation}
and then, since $\alpha$ and $\sigma$ are mute indexes, the
expression
\begin{equation}
{\frac {\delta {\cal H}} {d x^{\beta}}} = {\left(
\Gamma^{\alpha}_{\sigma \beta} - \Gamma^{\alpha}_{\sigma \beta}
\right)} {\left( {\frac {\partial {\cal H}} {\partial V_{\sigma}}}
\right)} {\left( {\frac {\partial {\cal H}} {\partial V^{\alpha}}}
\right)} \equiv 0, \label{s43}
\end{equation}
is identically equal to zero. In other words, ${\cal H}$ is
absolutely conserved, and this is very important since the analogy
with a canonical Hamiltonian structure is straightforward. In
fact, if, as above,
\begin{equation}
{\cal H} = {\cal H} {\left( p, q \right)} \label{s44}
\end{equation}
is a classical generic Hamiltonian function, expressed in the
canonical phase-space variables $\{p,q\}$, the total variation (in
a vector space $\mathbf{E_{2n}}$ whose dimensions are generically
given by $p_i$ and $q_j$ with $i, j = 1, ..., n$) is
\begin{equation}
d {\cal H} = {\frac {\partial {\cal H}} {\partial q}} d q + {\frac
{\partial {\cal H}} {\partial p}} d p, \label{s45}
\end{equation}
and
\begin{eqnarray}
{\frac {d {\cal H}} {dt}} &=& {\frac {\partial {\cal H}} {\partial
q}} \dot{q} + {\frac {\partial {\cal H}} {\partial p}} \dot{p} \nonumber\\
&=& {\frac {\partial {\cal H}} {\partial q}} {\frac {\partial
{\cal H}} {\partial p}} - {\frac {\partial {\cal H}} {\partial p}}
{\frac {\partial {\cal H}} {\partial q}} \equiv 0, \label{s47}
\end{eqnarray}
thanks to the Hamilton canonical equations
\begin{equation}
\dot{q} = {\frac {\partial {\cal H}} {\partial p}}, \qquad\qquad
\dot{p} = -{\frac {\partial {\cal H}} {\partial q}}. \label{s48}
\end{equation}
Such a canonical approach holds also in our covariant case if we
operate the substitutions
\begin{equation}
V^{\alpha} \, \longleftrightarrow \, p \qquad\qquad V_{\alpha} \,
\longleftrightarrow \, q \label{s49}
\end{equation}
and the canonical equations are
\begin{eqnarray}
{\frac {\delta V^{\alpha}} {d x^{\beta}}} = -
\Gamma^{\alpha}_{\sigma \beta} {\left( {\frac {\partial {\cal H}}
{\partial V_{\sigma}}} \right)} \quad & \longleftrightarrow &
\quad {\frac {d p} {d t}} = -{\frac {\partial {\cal H}} {\partial
q}}, \label{s50}\\ \  \ \nonumber\\
{\frac {\delta V_{\alpha}} {d x^{\beta}}} =
\Gamma^{\sigma}_{\alpha \beta} {\left( {\frac {\partial {\cal H}}
{\partial V^{\sigma}}} \right)} \quad & \longleftrightarrow &
\quad {\frac {d q} {d t}} = {\frac {\partial {\cal H}} {\partial
p}} \label{s51}.
\end{eqnarray}
In other words, starting from the (Hamiltonian) invariant
(\ref{s26}), we have recovered a covariant canonical symplectic
structure. The variation (\ref{s36}) may be seen as the generating
function ${\cal G}$ of canonical transformations where the
generators of $q-$, $p-$ and $t-$changes are dealt under the same
standard.

At this point, some important remarks have to be done. The
covariant and contravariant vector fields can be also of different
nature so that the above fundamental Hamiltonian invariant can be
generalized as
\begin{equation}
{\cal H} = W^{\alpha} V_{\alpha}, \label{inv1}
\end{equation}
or, considering scalar smooth and regular functions, as
\begin{equation}
{\cal H} = f {\left( W^{\alpha} V_{\alpha} \right)}, \label{inv2}
\end{equation}
or, in general
\begin{equation}
{\cal H} = f {\left( W^{\alpha} V_{\alpha}, B^{\alpha \beta}
C_{\alpha \beta}, B^{\alpha \beta} V_{\alpha} V^{\prime}_{\beta},
\dots \right)}, \label{bivector}
\end{equation}
where the invariant can be constructed by covariant vectors,
bivectors and tensors. Clearly, as above, the identifications
\begin{equation}
W^{\alpha} \, \longleftrightarrow \, p \qquad\qquad V_{\alpha} \,
\longleftrightarrow \, q \label{inv3}
\end{equation}
hold and the canonical equations are
\begin{equation}
{\frac {\delta W^{\alpha}} {d x^{\beta}}} = -
\Gamma^{\alpha}_{\sigma \beta} {\left( {\frac {\partial {\cal H}}
{\partial V_{\sigma}}} \right)} \qquad {\frac {\delta V_{\alpha}}
{d x^{\beta}}} = \Gamma^{\sigma}_{\alpha \beta} {\left( {\frac
{\partial {\cal H}} {\partial W^{\sigma}}} \right)}. \label{inv4}
\end{equation}
Finally, conservation laws are given by
\begin{equation}
{\frac {\delta {\cal H}} {d x^{\beta}}} = {\left(
\Gamma^{\alpha}_{\sigma \beta} - \Gamma^{\alpha}_{\sigma \beta}
\right)} {\left( \frac {\partial {\cal H}} {\partial V_{\sigma}}
\right)} {\left( \frac {\partial {\cal H}} {\partial W^{\alpha}}
\right)} \equiv 0. \label{hamiltongen}
\end{equation}
In our picture, this means that the canonical symplectic structure
is assigned in the way in which covariant and contravariant vector
fields are related. However, if the Hamiltonian invariant is
constructed by bivectors and tensors, equations (\ref{inv4}) and
(\ref{hamiltongen}) have to be generalized but the structure is
the same. It is worth noticing that we never used the metric field
but only connections in our derivations.

These considerations can be made independent of the reference
frame if we define a suitable system of unitary vectors by which
we can pass from holonomic to anholonomic description and
viceversa. We can define the reference frame on the event manifold
${\cal M}$ as vector fields $e_{(k)}$ in event space and dual
forms $e^{(k)}$ such that vector fields $e_{(k)}$ define an
orthogonal frame at each point and
\begin{equation}
e^{(k)} {\left( e_{(l)} \right)} = \delta^{(k)}_{(l)}. \label{01}
\end{equation}
If these vectors are unitary, in a Riemannian 4-spacetime are the
standard {\it vierbiens} \cite{landau}.

If we do not limit this definition of reference frame by
orthogonality, we can introduce a {\it coordinate reference frame}
$\left( \partial_{\alpha}, d s^{\alpha} \right)$ based on vector
fields tangent to line $x^{\alpha} = const$. Both reference frames
are linked by the relations
\begin{equation}
e_{(k)} = e^{\alpha}_{(k)} \partial_{\alpha}; \qquad e^{(k)} =
e_{\alpha}^{(k)} d x^{\alpha}. \label{02}
\end{equation}
From now on, Greek indices will indicate holonomic coordinates
while Latin indices between brackets, the anholonomic coordinates
({\it vierbien} indices in 4-spacetimes). We can prove the
existence of a reference frame using the orthogonalization
procedure at every point of spacetime. From the same procedure, we
get that coordinates of frame smoothly depend on the point. The
statement about the existence of a global reference frame follows
from this. A smooth field on time-like vectors of each frame
defines congruence of lines that are tangent to this field. We say
that each line is a world line of an observer or a {\it local
reference frame}. Therefore a reference frame is a set of local
reference frames. The {\it Lorentz transformation} can be defined
as a transformation of a reference frame
\begin{equation}
{x'}^{\alpha} = f{\left( x^0, x^1, x^2, x^3, \dots, x^n \right)},
\label{03}
\end{equation}
\begin{equation}
{e'}^{\alpha}_{(k)} = A^{\alpha}_{\beta} B^{(l)}_{(k)} e^{\beta}
_{(l)}, \label{04}
\end{equation}
where
\begin{equation}
A^{\alpha}_{\beta} = {\frac {\partial {x'}^{\alpha}} {\partial
{x'}^{\beta}}}, \qquad \delta_{(i)(l)} B^{(i)}_{(j)} B^{(l)}_{(k)}
= \delta_{(j)(k)}. \label{05}
\end{equation}
We call the transformation $A^{\alpha}_{\beta}$ the holonomic part
and transformation $B^{(l)}_{(k)}$ the anholonomic part.

A vector field $V$ has two types of coordinates: {\it holonomic
coordinates} $V^{\alpha}$ relative to a coordinate reference frame
and {\it anholonomic coordinates} $V^{(k)}$ relative to a
reference frame. For these two kinds of coordinates, the relation
\begin{equation}
V^{(k)} = e_{\alpha}^{(k)} V^{\alpha}\,, \label{06}
\end{equation}
holds. We can study parallel transport of vector fields using any
form of coordinates. Because equations (\ref{03}) and (\ref{04})
are linear transformations, we expect that parallel transport in
anholonomic coordinates has the same form as in holonomic
coordinates. Hence we write
\begin{equation}
D V^{\alpha} = d V^{\alpha} + \Gamma^{\alpha}_{\beta \gamma}
V^{\beta} d x^{\gamma}, \label{07}
\end{equation}
\begin{equation}
D V^{(k)} = d V^{(k)} + \Gamma^{(k)}_{(l)(p)} V^{(l)} d x^{(p)}.
\label{081}
\end{equation}
Because $D V^{\alpha}$ is also a tensor, we get
\begin{equation}
\Gamma^{(k)}_{(l)(p)} = e^{\alpha}_{(l)} e^{\beta}_{(p)}
e^{(k)}_{\gamma} \Gamma^{\gamma}_{\alpha \beta} + e^{\alpha}_{(l)}
e^{\beta}_{(p)} {\frac {\partial e^{(k)}_{\alpha}} {\partial
x^{\beta}}}. \label{08}
\end{equation}
Eq.(\ref{08}) shows the similarity between holonomic and
anholonomic coordinates. Let us  introduce the symbol
$\partial_{(k)}$ for the derivative along the vector field
$e_{(k)}$
\begin{equation}
\partial_{(k)} = e^{\alpha}_{(k)} \partial_{\alpha}. \label{09}
\end{equation}
Then Eq.(\ref{08}) takes the form
\begin{equation}
\Gamma^{(k)}_{(l)(p)} = e^{\alpha}_{(l)} e^{\beta}_{(p)}
e^{(k)}_{\gamma} \Gamma^{\gamma}_{\alpha \beta} + e^{\alpha}_{(l)}
\partial_{(p)} e^{(k)}_{\alpha}. \label{010}
\end{equation}
Therefore, when we move from holonomic coordinates to anholonomic
ones, the connection also transforms the way similarly to when we
move from one coordinate system to another. This leads us to the
model of anholonomic coordinates. The vector field $e_{(k)}$
generates lines defined by the differential equations
\begin{equation}
e^{\alpha}_{(l)} {\frac {\partial \tau} {\partial x^{\alpha}}} =
\delta^{(k)}_{(l)}, \label{011}
\end{equation}
or the symbolic system
\begin{equation}
{\frac {\partial \tau} {\partial x^{(l)}}} = \delta^{(k)}_{(l)}.
\label{012}
\end{equation}
Keeping in mind the symbolic system (\ref{012}), we denote the
functional $\tau$ as $x^{(k)}$ and call it the anholonomic
coordinate. We call the regular coordinate holonomic. Then we can
find derivatives and get
\begin{equation}
{\frac {\partial x^{(k)}} {\partial x^{\alpha}}} =
\delta^{(k)}_{\alpha}. \label{013}
\end{equation}
The necessary and sufficient conditions to complete the
integrability of system (\ref{013}) are
\begin{equation}
\omega^{(i)}_{(k)(l)} = e^{\alpha}_{(k)} e^{\beta}_{(l)} {\left(
{\frac {\partial e^{(i)}_{\alpha}} {\partial x^{\beta}}} - {\frac
{\partial e^{(i)}_{\beta}} {\partial x^{\alpha}}} \right)} = 0,
\label{014}
\end{equation}
where we introduced the anholonomic object
$\omega^{(i)}_{(k)(l)}$. Therefore each reference frame has $n$
vector fields
\begin{equation}
\partial_{(k)} = {\frac {\partial} {\partial x^{(k)}}} =
e^{\alpha}_{(k)} \partial_{\alpha}, \label{015}
\end{equation}
which have the commutators
\begin{eqnarray}
{\left[ \partial_{(i)}, \partial_{(j)} \right]} = {\left(
e^{\alpha}_{(i)} \partial_{\alpha} e^{\beta}_{(j)} -
e^{\alpha}_{(j)} \partial_{\alpha} e^\beta_{(i)} \right)}
e^{(m)}_{\beta} \partial_{(m)} \nonumber \\
= e^{\alpha}_{(i)} e^{\beta}_{(j)} {\left( - \partial_{\alpha}
e^{(m)}_{\beta} + \partial_{\beta} e^{(m)}_{\beta} \right)}
\partial_{(m)} = \omega^{(m)}_{(i)(j)}
\partial_{(m)}. \label{016}
\end{eqnarray}
For the same reason, we introduce the forms
\begin{equation}
dx^{(k)} = e^{(k)} = e_{\beta}^{(k)} d x^{\beta}, \label{017}
\end{equation}
and a differential of this form is
\begin{eqnarray}
d^2 x^{(k)} = d {\left( e^{(k)}_{\alpha} d x^{\alpha} \right)} =
{\left( \partial_{\beta} e^{(k)}_{\alpha} - \partial_{\alpha}
e^{(k)}_{\beta} \right)} d x^{\alpha} \wedge d x^{\beta} \nonumber
\\
= - \omega^{(m)}_{(k)(l)} d x^{(k)} \wedge dx^{(l)}. \label{018}
\end{eqnarray}
Therefore when $\omega^{(i)}_{(k)(l)} \neq 0$, the differential $d
x^{(k)}$ is not an exact differential and the system (\ref{013}),
in general, cannot be integrated. However, we can consider
meaningful objects which model the solution. We can study how the
functions $x^{(i)}$ changes along different lines. The functions
$x^{(i)}$ is a natural parameter along a flow line of vector
fields $e_{(i)}$. It is defined along any line.

All the above results  can be immediately achieved in holonomic
and anholonomic formalism considering the equation
\begin{equation}
{\cal H} = W^{\alpha} V_{\alpha} = W^{(k)} V_{(k)}, \label{989}
\end{equation}
and the analogous ones. This means that the results are
independent of the reference frame and the symplectic covariant
structure always holds.

\section{The hydrodynamic picture }
In order to further check the validity of the above approach, we
can prove that it is always consistent with the hydrodinamic
picture (see also \cite{ferapontov} for details on hydrodynamic
covariant formalism).

Let us  define a phase space density $f{\left( x, p; s \right)}$
which evolves according to the Liouville equation
\begin{equation}
{\frac {\partial f} {\partial s}} + {\frac {1} {m}} {\frac
{\partial} {\partial x^{\mu}}} {\left( g^{\mu \nu} p_{\nu} f
\right)} - {\frac {1} {2m}} {\frac {\partial} {\partial
p_{\lambda}}} {\left( {\frac {\partial g^{\mu \nu}} {\partial
x^{\lambda}}} p_{\mu} p_{\nu} f \right)} = 0. \label{Liouveq}
\end{equation}
Next we define the density $\varrho {\left( x; s \right)}$, the
covariant current velocity $v_{\mu} {\left( x; s \right)}$ and the
covariant stress tensor ${\cal P}_{\mu \nu} {\left( x; s \right)}$
according to the relations
\begin{equation}
\varrho {\left( x; s \right)} = m n \int {\rm d}^4 p f{\left( x,
p; s \right)}, \label{Density}
\end{equation}
\begin{equation}
\varrho {\left( x; s \right)} v_{\mu} {\left( x; s \right)} = n
\int {\rm d}^4 p p_{\mu} f{\left( x, p; s \right)},
\label{Velocity}
\end{equation}
\begin{equation}
{\cal P}_{\mu \nu} {\left( x; s \right)} = {\frac {n} {m}} \int
{\rm d}^4 p p_{\mu} p_{\nu} f{\left( x, p; s \right)}.
\label{Stress}
\end{equation}
It can be verified , by direct substitution, that a solution to
the Liouville Eq.(\ref{Liouveq}) of the form
\begin{equation}
f{\left( x, p; s \right)} = {\frac {1} {mn}} \varrho {\left( x; s
\right)} \delta^4 {\left[ p_{\mu} - m v_{\mu} {\left( x; s
\right)} \right]}, \label{Deltasol}
\end{equation}
leads to the equation of continuity
\begin{equation}
{\frac {\partial \varrho} {\partial s}} + {\frac {\partial}
{\partial x^{\mu}}} {\left( g^{\mu \nu} v_{\nu} \varrho \right)} =
0, \label{Continuity}
\end{equation}
and to the equation for balance of momentum
\begin{equation}
{\frac {\partial} {\partial s}} {\left( \varrho v_{\mu} \right)} +
{\frac {\partial} {\partial x^{\lambda}}} {\left( g^{\lambda
\alpha} {\cal P}_{\alpha \mu} \right)} + {\frac {1} {2}} {\frac
{\partial g^{\alpha \beta}} {\partial x^{\mu}}} {\cal P}_{\alpha
\beta} = 0. \label{Mombalance}
\end{equation}
Taking into account the fact that for the particular solution
(\ref{Deltasol}), the stress tensor, as defined by
Eq.(\ref{Stress}), is given by the expression
\begin{equation}
{\cal P}_{\mu \nu} {\left( x; s \right)} = \varrho v_{\mu}
v_{\nu}, \label{Stress1}
\end{equation}
we obtain the final form of the hydrodynamic equations
\begin{equation}
{\frac {\partial \varrho} {\partial s}} + {\frac {\partial}
{\partial x^{\mu}}} {\left( \varrho v^{\mu} \right)} = 0,
\label{Continuity1}
\end{equation}
\begin{equation}
{\frac {\partial v_{\mu}} {\partial s}} + v^{\lambda} {\left(
{\frac {\partial v_{\mu}} {\partial x^{\lambda}}} -
\Gamma^{\nu}_{\mu \lambda} v_{\nu} \right)} = {\frac {\partial
v_{\mu}} {\partial s}} + v^{\lambda} \nabla_{\lambda} v_{\mu} = 0.
\label{Mombalance1}
\end{equation}
It is straightforward to see that, through the substitution
$v_{\mu}\rightarrow V_{\mu}$, Eq.(\ref{s41})  is immediately
recovered along a geodesic, that is our covariant symplectic
structure is consistent with a hydrodynamic picture. It is worth
noticing that if ${\displaystyle \frac{\partial
v_{\mu}}{\partial{s}}}$ in Eq.(\ref{Mombalance1}), the motion is
not geodesic. The meaning of this term different from zero is that
an extra force is acting on the system.

\section{Applications, Discussion and Conclusions}
Several applications of the previous results can be achieved
specifying the nature of vector (or tensor) fields which define
the Hamiltonian conserved invariant ${\cal H}$. Considerations in
General Relativity and Electromagnetism are particularly
interesting at this point. Let us take into account the Riemann
tensor $ R^{\rho}_{\sigma \mu \nu}$. It comes out when a given
vector $V^{\rho}$ is transported along a closed path on a generic
curved manifold. It is
\begin{equation}
[\nabla_{\mu}, \nabla_{\nu}] V^{\rho} = R^{\rho}_{\sigma \mu \nu}
V^{\sigma}, \label{s13}
\end{equation}
where $\nabla_{\mu}$ is the covariant derivative. We are assuming
a Riemannian $\mathbf{V}_n$ manifold as standard in General
Relativity. If connection is not symmetric, an additive torsion
field comes out from the parallel transport.

Clearly, the Riemann tensor results from the commutation of
covariant derivatives and it can be expressed as the sum of two
commutators
\begin{equation}
R^{\rho}_{\sigma \mu \nu} = \partial_{[ \mu}, \Gamma^{\rho}_{\nu]
\sigma} + \Gamma^{\rho}_{\lambda [\mu}, \Gamma^{\lambda}_{\nu]
\sigma}. \label{s19}
\end{equation}
Furthermore, (anti) commutation relations and  cyclic identities
(in particular Bianchi's identities) hold for the Riemann tensor
\cite{landau}.

All these straightforward considerations suggest the presence of a
symplectic structure whose elements are covariant and
contravariant vector fields, $V^{\alpha}$ and $V_{\alpha}$,
satisfying the properties (\ref{s3})-(\ref{s6'}). In this case,
the dimensions of vector space $\mathbf{E}_{2n}$ are assigned by
$V^{\alpha}$ and $V_{\alpha}$. It is important to notice that such
properties imply the connections (Christoffel symbols) and not the
metric tensor.

As we said, the invariant (\ref{s26}) is a generic conserved
quantity specified by the choice of $V^{\alpha}$ and $V_{\alpha}$.
If
\begin{equation}
V^{\alpha} = {\frac {d x^{\alpha}} {d s}},
\end{equation}
is a 4-velocity, with $\alpha = 0, 1, 2, 3$, immediately, from
Eq.(\ref{s50}), we obtain the equation of geodesics of General
Relativity,
\begin{equation}
{\frac {d^2 x^{\alpha}} {d s^2}} + \Gamma^{\alpha}_{\mu \nu}
{\frac {d x^{\mu}} {d s}} {\frac {d x^{\nu}} {d s}} = 0.
\label{s52}
\end{equation}
On the other hand, being
\begin{equation}
\delta V^{\alpha} = R^{\alpha}_{\beta \mu \nu} V^{\beta} d
x^{\mu}_{1} d x^{\nu}_{2}, \label{s53}
\end{equation}
the result of the transport along a closed path, it is easy to
recover the geodesic deviation considering the geodesic
(\ref{s52}) and the infinitesimal variation $\xi^{\alpha}$ with
respect to it, i.e.
\begin{equation}
{\frac {d^2 {\left( x^{\alpha} + \xi^{\alpha} \right)}} {d s^2}} +
\Gamma^{\alpha}_{\mu \nu} (x + \xi) {\frac {d {\left( x^{\mu} +
\xi^{\mu} \right)}} {d s}} {\frac {d {\left( x^{\nu} + \xi^{\nu}
\right)}} {ds}} = 0, \label{s55}
\end{equation}
which gives, through Eq.(\ref{s19}),
\begin{equation}
{\frac {d^2 \xi^{\alpha}} {d s^2}} = R^{\alpha}_{\mu \lambda \nu}
{\frac {d x^{\mu}} {ds}} {\frac {dx^{\nu}} {ds}} \xi^{\lambda}.
\label{s56}
\end{equation}
Clearly the symplectic structure is due to the fact that the
Riemann tensor is derived from covariant derivatives either as
\begin{equation}
{\left[ \nabla_{\mu}, \nabla_{\nu} \right]} V^{\rho} =
R^{\rho}_{\sigma \mu \nu} V^{\sigma}, \label{s133}
\end{equation}
or
\begin{equation}
{\left[ \nabla_{\mu}, \nabla_{\nu} \right]} V_{\rho} =
R^{\sigma}_{\mu \nu \rho} V_{\sigma}. \label{s1333}
\end{equation}
In other words, fundamental equations of General Relativity are
recovered from our covariant symplectic formalism.

Another  interesting choice allows  to recover the standard
Electromagnetism. If $V^{\alpha} = A^{\alpha}$, where $A^{\alpha}$
is the vector potential  and the Hamiltonian invariant is
\begin{equation}
{\cal H} = A^{\alpha} A_{\alpha}\,, \label{s58}
\end{equation}
it is straightforward, following the above procedure, to obtain,
from the covariant Hamilton equations, the electromagnetic tensor
field
\begin{equation}
F_{\alpha \beta} = \nabla_{\alpha} A_{\beta} - \nabla_{\beta}
A_{\alpha} = \nabla_{[\alpha} A_{\beta]}, \label{s59}
\end{equation}
and the Maxwell equations (in a generic empty curved spacetime)
\begin{equation}
\nabla^{\alpha} F_{\alpha \beta} =0, \qquad
\nabla_{[\alpha}F_{\lambda \beta]} = 0. \label{s60}
\end{equation}
The standard Lorentz gauge is
\begin{equation}
\nabla^{\alpha} A_{\alpha} = 0, \label{s61}
\end{equation}
and electromagnetic wave equation is easily recovered.

In summary,  a covariant, symplectic structure can be found for
every Hamiltonian invariant which can be constructed by covariant
vectors, bivectors and tensor fields. In fact,  any theory of
physics has to be endowed with a symplectic structure in order to
be formulated at a fundamental level.

We pointed out that curvature invariants of General Relativity can
show such a feature and, furthermore, they can be recovered from
Hamiltonian invariants opportunely defined. Another interesting
remark deserves the fact that, starting from such invariants,
covariant and contravariant vector fields can be read as the
configurations $q^i$ and the momenta $p_i$ of classical
Hamiltonian dynamics so then the Hamilton-like equations of motion
are recovered from the application of covariant derivative to both
these vector fields. Besides, the approach can be formulated in a
holonomic and anholonomic representations, once vector fields (or
tensors in general) are represented in {\it vierbien} or
coordinate--frames. This feature is essential to be sure that
general covariance and symplectic structure are conserved in any
case.

 Specifying the nature of vector fields, we select the
particular theory. For example, if the vector field is the
4-velocity, we obtain geodesic motion and geodesic deviation. If
the vector is the vector potential of Electromagnetism, Maxwell
equations and Lorentz gauge are recovered. The scheme is
independent of the nature of vector field and, in our opinion, it
is a strong hint toward a unifying view of basic interactions,
gravity included.

\begin{acknowledgments}
We wish to acknowledge Francesco Guerra for the useful discussions
and suggestions on the topic.
\end{acknowledgments}

\vfill

\end{document}